\documentclass[preprint,12pt]{aastex}

\usepackage{latexsym}
\newcommand {\ha}{H$\alpha$}

\newcommand {\hii}{H\,{\sc ii}}
\newcommand {\kms}{km~s$^{-1}$}

\newcommand {\oi}{[O\,{\sc i}]}
\newcommand {\oiii}{[O\,{\sc iii}]}
\newcommand {\nii}{[N\,{\sc ii}]}
\newcommand {\sii}{[S\,{\sc ii}]}

\shorttitle{Ultraluminous SNR in NGC\,6946}
\shortauthors{Dunne, Gruendl, \& Chu}
\slugcomment{To appear in The Astronomical Journal, March 2000}

\begin{document}

\title{What Produced the Ultraluminous Supernova Remnant in NGC\,6946?}

\author{Bryan C. Dunne, Robert A. Gruendl, and You-Hua Chu}
\affil{Department of Astronomy, 
University of Illinois,
1002 West Green Street, 
Urbana, IL 61801}
\email{carolan@astro.uiuc.edu}

\begin{abstract}
The ultraluminous supernova remnant (SNR) in NGC\,6946 
is the brightest known SNR in X-rays, $\sim$1000 times
brighter than Cas A.  To probe the nature of this remnant and
its progenitor, we have obtained high-dispersion optical echelle 
spectra.  The echelle spectra detect \ha, \nii, and \oiii\ lines,
and resolve these lines into a narrow (FWHM $\sim$20--40~\kms) 
component from un-shocked material and a broad (FWHM $\sim$250~\kms) 
component from shocked material.  Both narrow and broad 
components have unusually high \nii/\ha\ ratios, $\sim$1.
Using the echelle observation, archival {\it HST} images, and
archival {\it ROSAT} X-ray observations, we conclude that the 
SNR was produced by a normal supernova, whose progenitor 
was a massive star, either a WN star or a luminous blue variable.
The high luminosity of the remnant is caused by the supernova ejecta
expanding into a dense, nitrogen-rich circumstellar nebula
created by the progenitor.
\end{abstract}


\clearpage

\section{Introduction}

NGC\,6946 is a face-on Sc spiral galaxy at a distance of 5.1~Mpc 
\citep{dV79}.  Twenty-seven supernova remnant (SNR) 
candidates have been reported in this galaxy by \citet{MF97}, 
using \ha\ and \sii\ images.  Because of the large 
distance to NGC\,6946 and limited sensitivity of X-ray and radio
detectors, it is difficult to confirm these SNR candidates.  The 
most luminous SNR candidate in NGC\,6946, listed as MF16 in \citet{MF97}, 
is the only one of the 27 candidates confirmed with radio and X-ray 
observations \citep{vDyk94,S94}.

MF16, at \mbox{$\alpha$$=$20$^{\rm h}$35$^{\rm m}$0\fs8}, 
\mbox{$\delta$$=$$+$60\arcdeg11\arcmin30\farcs5} (J2000), 
was the first known SNR in NGC\,6946 because of its extremely 
high luminosities in both the X-ray and optical bands \citep{S94,BF94}.  
It is the brightest known SNR in X-rays, $\sim$1000 
times as luminous as the young Galactic remnant Cas A  
and $\sim$3.5 times as luminous as the young, ultraluminous SNR in 
NGC\,4449 \citep{BFS97,BF94}.  Its optical brightness is roughly the 
same as that of the NGC\,4449 SNR and over 10 times greater than 
that of N49 \citep{BF94}, the brightest SNR in the 
Large Magellanic Cloud (LMC).  
Such high luminosities are usually associated with young 
remnants, as is high velocity gas.  Yet, gas with 
\mbox{V$_{\rm exp} > 600$ \kms} has not been detected
from MF16.  {\it Hubble Space Telescope} ({\it HST}) 
images of this remnant show multiple 
loops.  This has led \citet{BFS99} to suggest
that MF16 actually consists of colliding SNRs of different 
ages.

Intrigued by the extremely high luminosity and interesting morphology,
we have obtained a high-dispersion echelle spectrum of MF16 to 
investigate its physical nature and the cause of its very
high luminosity.  The echelle spectrum clearly resolves the emission
lines into a broad component and a narrow component, which presumably
correspond to the shocked and un-shocked material, respectively.
From the velocity profile of the \ha\ line and the {\it HST} 
F656N image, we are able to derive the mass and kinetic energy
of the SNR.  The echelle spectrum has also detected multiple
nebular lines, allowing us to examine the diagnostic line ratios.
We find that the \nii$\lambda$6584 to \ha\ line ratio is anomalously 
high for both the narrow and broad components.  The \nii$\lambda$6584/\ha\ 
ratio of the narrow line component is much higher than those 
of \hii\ regions in NGC\,6946 at similar galactrocentric 
distances.  A high \nii/\ha\ ratio is frequently seen in nebulae 
around Wolf-Rayet stars and luminous blue variables; for 
all such nebulae with good temperature measurements, an 
enhanced nitrogen abundance is found (e.g., Esteban et al. 1992;
Smith et al. 1998).  The high \nii$\lambda$6584/\ha\ ratio in
the narrow component of the SNR MF16 is thus a significant 
discovery.  To determine how unusual this high \nii$\lambda$6584/\ha\ ratio
is, we have compared MF16 to other SNR candidates in NGC\,6946.
We have also examined echelle spectra of eight SNRs in M33,
and compared the \nii$\lambda$6584/\ha\ ratios for both the narrow and broad 
components.  We have further explored the nature of MF16 by 
examining its X-ray spectra and comparing it to several SNRs 
in the LMC.

In this paper, we report on our observations and the results of our
analysis of the ultraluminous SNR MF16 in NGC\,6946.
The details of the datasets are described in \S~\ref{sec:data}, 
our analysis is presented in \S~\ref{sec:analysis}, and the 
results are discussed in \S~\ref{sec:discuss}.

\section{Observations and Archival Datasets}
\label{sec:data}

The data we used to study the SNR MF16 in NGC\,6946 included 
available archived {\it Hubble Space Telescope} ({\it HST}) 
observations, high-dispersion echelle spectra, and archived {\it ROSAT}
observations.  For comparison, 
optical echelle spectra of a control group of eight 
SNRs in M33 and archived {\it ROSAT} observations of three SNRs in
the LMC were also examined.  All of the data were reduced using 
standard routines in IRAF\footnote{Image Analysis and Reduction
Facility -- IRAF is distributed by the 
National Optical Astronomy Observatories, which are operated by 
the Association of Universities for Research in Astronomy, Inc., 
under cooperative agreement with the National Science Foundation.} 
and the STSDAS\footnote{Space Telescope Science Data Analysis 
System -- http://ra.stsci.edu/} and PROS\footnote{PROS/XRAY
Data Analysis System -- http://hea-www.harvard.edu/PROS/pros.html} 
packages.

\subsection{{\it HST} WFPC2 Images}
\label{sec:HST}

{\it HST} Wide Field/Planetary Camera 2 (WFPC2) observations of 
MF16 (PI: Blair) 
took place on 1996 January 27 (UT).  The remnant was centered on
the PC chip for all exposures.  A 2$\times$700~s observation was
made with the F656N filter (\ha) using CR-split to improve 
cosmic ray rejection.  A 2$\times$400~s observation was
made with the F439W filter (blue continuum) again using CR-split.
A 700~s observation was made with the F555W filter (visual continuum);
CR-split was not used for this observation, making rigorous 
cosmic-ray removal difficult.
The PC has a plate scale of 0\farcs0455~pixel$^{-1}$ 
over a 36\arcsec\ field-of-view.  The raw data were recalibrated 
using the most recent reference files recommended by 
the Space Telescope Science Institute.   
The flux calibration was obtained using the methods outlined 
on the WFPC2 website\footnote{http://www.stsci.edu/intrsuments/wfpc2/}.
Images from the WFPC2 observations showing MF16 and neighboring regions
are presented in Figure~\ref{figHST}.   

\subsection{Echelle Spectra}
\label{sec:echelle}

\begin{figure}
\epsscale{0.64}
\plotone{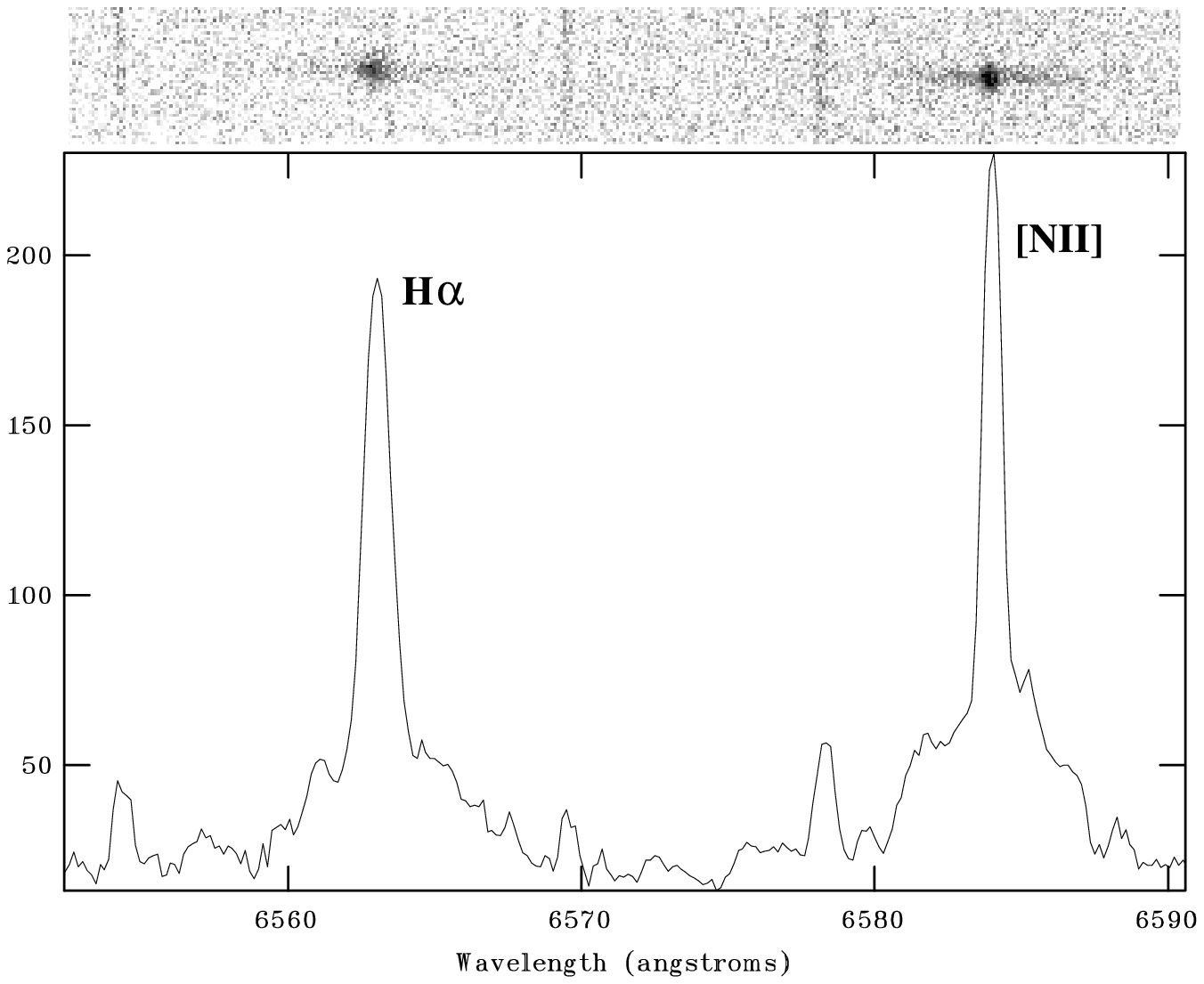}
\caption{Echellogram and corresponding spectrum of MF16 
with \ha\ and \nii\ lines shown.  Both a narrow and broad line component 
of the emission from the SNR are evident. \label{figechelle}}
\end{figure}

Our high-dispersion spectra of MF16 were 
obtained using the echelle spectrograph on the 4-m telescope 
at Kitt Peak National Observatory (KPNO) on 1999 March 4 (UT).
The \mbox{79 l mm$^{-1}$} echelle grating was used in combination with
a \mbox{226 l mm$^{-1}$} cross disperser and the long focus red camera to 
achieve a reciprocal dispersion of \mbox{3.5 \AA\ mm$^{-1}$} at \ha.  
The spectra were imaged with the T2KB CCD detector.  The pixel
size is 24~$\mu$m, corresponding to 0\farcs24~pixel$^{-1}$ along
the slit and \mbox{$\sim$3.7 \kms\ pixel$^{-1}$} along the 
dispersion axis.  The SNR MF16 was observed for 600~s with 
a east--west oriented 15\arcsec-long slit of width 2\arcsec.  The 
remnant was contained entirely within the slit.  Despite
the short exposure time, the sky lines are prominent and thus
can be used to determine the instrumental FWHM, 17$\pm$1~\kms,
and to fine-tune the velocity calibration.  The \ha--\nii$\lambda$6584 region
of the echellogram is presented in Figure~\ref{figechelle}.

\begin{figure}
\epsscale{0.64}
\plotone{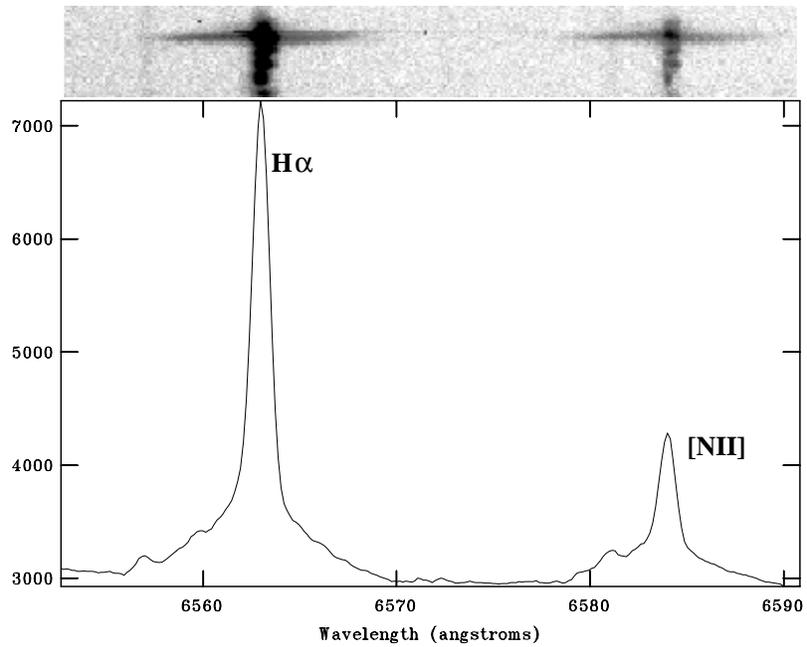}
\caption{Echellogram and corresponding spectrum of M33-8 
with \ha\ and \nii\ lines shown.  As in Figure~\ref{figechelle},
both a narrow and broad line component of the emission from the 
SNR are evident. \label{figm33}}
\end{figure}

The high-dispersion spectra of the SNRs in M33 were
also obtained using the echelle spectrograph on the 4-m telescope
at KPNO, but on 1986 September 16--17 (UT).
Echellograms for M33-2, M33-4, M33-6, M33-8, M33-9, M33-11,
M33-16, and M33-18 (notation from D'Odorico, Dopita, \& Benvenuti
1980) were obtained over the two nights.
The \mbox{79 l mm$^{-1}$} echelle grating was used.  A flat mirror,
replacing the cross disperser, and a broad \ha\ post-slit filter
were used to isolate a single order around the \ha\ line.
These in combination with the UV camera 
achieved a reciprocal dispersion of \mbox{8.0 \AA\ mm$^{-1}$} at \ha.
The spectra were imaged with the TI-4 CCD detector.  The pixel
size was 15~$\mu$m, corresponding to 0\farcs27~pixel$^{-1}$ along
the slit and \mbox{$\sim$5.5 \kms\ pixel$^{-1}$} along the
dispersion axis.  The M33 SNRs were observed with a east--west
oriented long slit of width 3\farcs3 for 1,800~s (M33-16),
2,400~s (M33-7), 2,700~s (M33-8, M33-9, M33-18),
3,000~s (M33-6), and 3,600~s (M33-2, M33-4, and M33-11).  
Th-Ar lamp comparison spectra were used to 
determine the instrumental FWHM, 31$\pm$1~\kms.
As an example of the echelle spectra obtained, the \ha--\nii\
region of M33-8's echellogram is presented in Figure~\ref{figm33}.

\subsection{{\it ROSAT} PSPC Observations}
\label{sec:rosat}

{\it ROSAT} observations of the galaxy NGC\,6946 
(PI: Schlegel) were made on 1992 June 11 (UT).  The 
galaxy was observed for 36,713~s with the Position 
Sensitive Proportional Counter (PSPC).  The sequence 
number of the observation is rp600272.  A complete
description of the observation can be found in \citet{S94}.
For comparison, we have also used {\it ROSAT} PSPC 
observations of LMC SNRs N49 (rp500062; 5,870~s; PI: Fink), 
N132D (rp500141; 6,212~s; PI: Hughes), 
and N157B (rp500131; 16,069~s; PI: Chu).  The X-ray
spectra are presented in Figure~\ref{figxray}.

\section{Results and Analysis}
\label{sec:analysis}

\subsection{Stellar and Interstellar Properties}
\label{sec:environment}

\begin{figure}
\epsscale{1}
\plotone{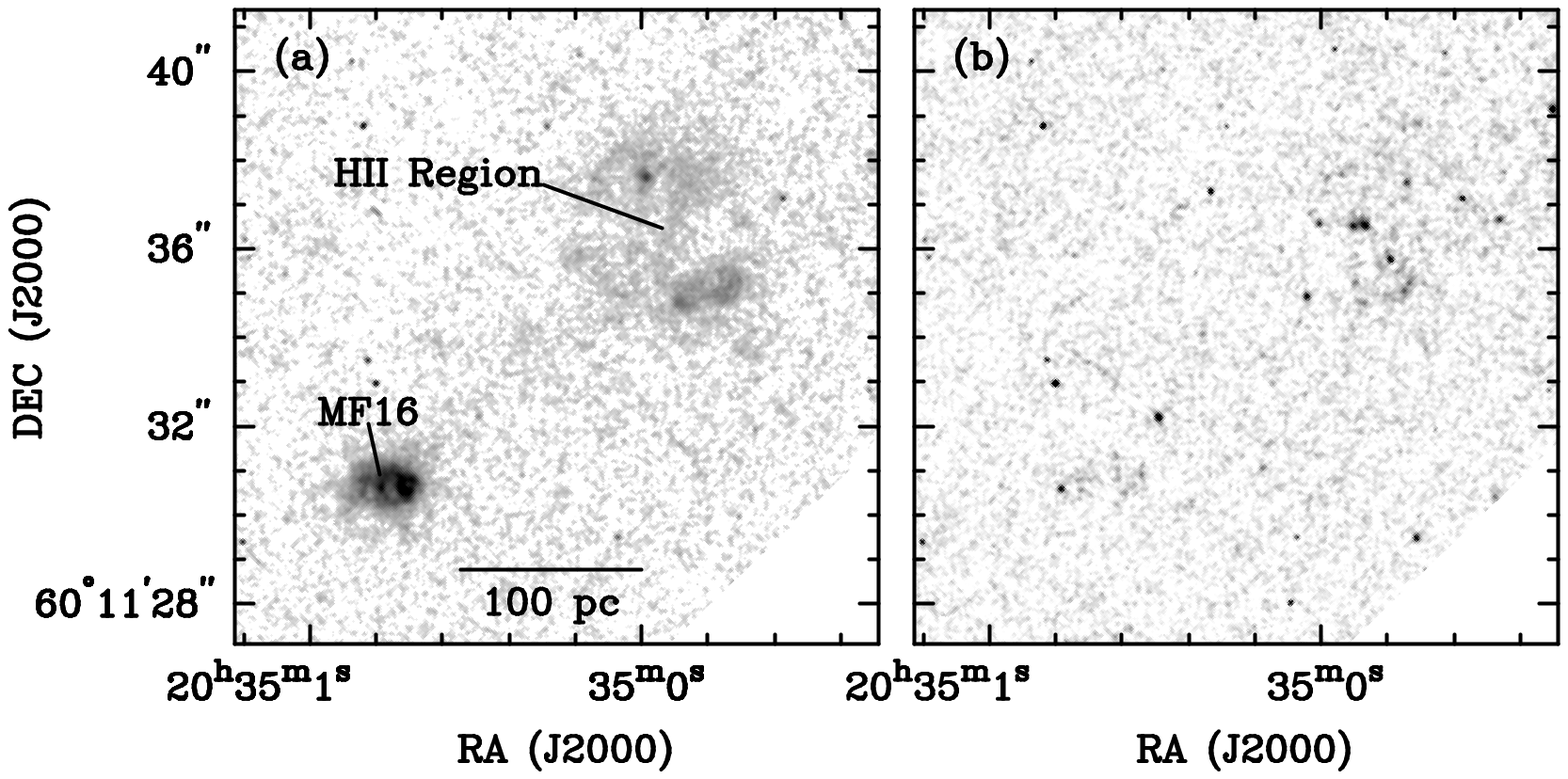}
\caption{WFPC2 images of MF16 and surrounding region: 
(a) F656N image (\ha), (b) F439W image (blue continuum).  The location 
of MF16 and a nearby \hii\ region are indicated in (a). \label{figHST}}
\end{figure}

We have used the WFPC2 images to 
measure many of the basic properties of MF16.
As measured using the F656N image (See Figure~\ref{figHST}a), 
MF16 has an angular size of \mbox{0\farcs8 $\times$ 1\farcs2},
corresponding to \mbox{20 pc $\times$ 30 pc} at a distance
of 5.1~Mpc.  In the F439W image (See Figure~\ref{figHST}b), 
only a single continuum point source 
is projected within the remnant.  To determine the origin of 
this point source, we first measured its apparent magnitude, 
\mbox{m$_{\rm F439W}$=22.7$\pm$0.2 mag}, using the ``phot'' package 
in IRAF.  We then applied a correction of 0.66~mag given 
by \citet{WFPC2} to convert to m$_{\rm B}$.  Using the 
\mbox{E(B$-$V)=0.52 mag} from \citet{BF94} and a distance 
of 5.1~Mpc, we have determined that \mbox{M$_{\rm B}$=$-$7.3 mag} 
for the point source\footnote{The F555W image is contaminaed
by nebular \oiii\ emission from MF16, so the brightness of
the point source cannot be accurately measured and visual
magnitude cannot be derived.}.  The point source has a FWHM of 
$\sim$2~pixels, and therefore it could be up to 
$\sim$2.5~pc across.  This size is too small for the 
point source to be an open cluster or OB association 
\citep{MB81,LH70}.  The luminosity of the point 
source is consistent with a single class Ia supergiant or a
multiple star system with fainter individual stars
\citep{SK82}.  

We have also explored the stellar and interstellar 
environments surrounding MF16.  The remnant is in a 
rather unpopulated region of NGC\,6946, many hundreds 
of parsecs north of a spiral arm.  The F656N image 
shows that MF16 is not associated with any obvious 
\hii\ region (See Figure~\ref{figHST}a).  A few faint point 
sources roughly 1\arcsec\ west of MF16 are visible in the 
WFPC2 F555W and F439W images (See Figure~\ref{figHST}b).  
To determine whether these point sources constitute an OB association
we have made photometric measurements.  These measurements
have high uncertainties because the sources are faint; the typical 
uncertainty is 0.5 mag in m$_{\rm F439W}$ and 0.2 mag in m$_{\rm F555W}$.  
If the extinction is negligible, the spectral types and luminosity
classes of these stars range between B--early-A main-sequence stars and 
F--G class II bright giants \citep{SK82}.  These spectral types are too late 
for this group of stars to be an OB association.  On the other hand, 
due to the large uncertainities in the magnitude measurements,
and hence colors, these stars could be brighter, have earlier
spectral types, and may belong to an OB association.  Nevertheless, 
the lack of an associated \hii\ region indicates that either
these stars have no ionizing power or there is no gas in the
surrounding region; the extinction is therefore unknown.
In either case, the association must be old, at
least $\sim$10$^{7}$~yr old.

The closest large structure to MF16 is an \hii\ region
centered $\sim$200~pc to the northwest of the remnant at 
\mbox{$\alpha$$=$20$^{\rm h}$34$^{\rm m}$59\fs9}, 
\mbox{$\delta$$=$$+$60\arcdeg11\arcmin36\farcs3} (J2000).  
The location of the \hii\ region is indicated on the
F656N image (See Figure~\ref{figHST}a).  The F439W image 
also shows a stellar concentration in this same location
(See Figure~\ref{figHST}b).  The morphology and spatial 
distribution of this \hii\ region/stellar concentration are 
qualitatively similar to some \hii\ regions/OB associations 
in the LMC \citep{LH70}.
The stellar concentration is likely an OB association roughly
75~pc across.  This size is similar to the average OB
association sizes for the LMC, $\sim$70~pc \citep{LH70}, 
and M31, $\sim$90~pc \citep{Mag93}.  \citet{BFS99} have
raised the possibility that MF16 is associated with
this stellar concentration.  To include the stellar 
concentration and the stellar sources in and around MF16, 
the OB association would be $\sim$250~pc across, far larger 
than the normal range indicated by LMC and M31 OB associations
\citep{LH70,Mag93}.  The OB association would also be required
to extend over a region of space $\sim$120~pc
across that does not appear to contain a strong 
concentration of bright stars in the WFPC2 F439W image.  
This requirement is not compatible with the results of 
either older ``by eye'' OB association searches, such 
as \citet{LH70}, or recent automated OB association 
searches using the Path Linkage Criterion or the 
``Friends of Friends'' algorithm \citep{Bat91,Wil91}.
MF16 is therefore unlikely to be associated with the stellar
concentration.

\subsection{Kinematic and Spectral Properties}
\label{sec:kinematic}

The echelle observation of the SNR MF16 detected the \ha, 
\nii$\lambda$$\lambda$6548, 6584, \oiii$\lambda$$\lambda$4959, 5007, 
and \oi$\lambda$6300 lines. The \sii$\lambda$$\lambda$6716, 
6731 lines are not detected as they were off the echelle
blaze.  The \ha, \nii$\lambda$6584, and \oiii$\lambda$5007 
lines show a narrow 
component superposed on a broad component.
The \nii$\lambda$6548, \oiii$\lambda$4959 and \oi$\lambda$6300 lines
are too weak to detect the broad line component.
We have determined the velocity widths of the broad 
and narrow components for the well-detected lines.
The narrow components typically have FWHM of 20--40~\kms,
while the broad components have FWHM $\sim$250~\kms\
(measured only for \ha\ and \nii) and FWZI $\sim$425~\kms\
(See Table~\ref{tblwidth}).  We have taken the expansion 
velocity of the remnant to be the half-width at zero 
intensity of the \ha\ broad component and obtained 
\mbox{V$_{\rm exp}$ $\approx$ 225~\kms}.  

The majority of the \ha, \nii$\lambda$6584, and \oiii$\lambda$5007 
emission from MF16 comes from the broad component of the lines.  
In \ha, the broad component is $\sim$1.5 times as strong as 
the narrow component.  In \nii$\lambda$6584, the broad component 
is $\sim$1.8 times as strong as the narrow component.  
In \oiii$\lambda$5007, the broad component is only slightly stronger 
than the narrow component.
The most striking and unusual feature of the
spectra is that the total \nii$\lambda$6584 flux is comparable to 
the \ha\ flux.  The \nii$\lambda$6584/\ha\ ratio of the broad 
component is near unity while that of the narrow 
component is slightly less.  A full detailing of 
comparative flux levels can be found in Table~\ref{tblflux}.  

\subsection{Physical Properties}
\label{sec:phys}

We have measured a total flux for MF16 of $\sim$5,000 counts 
from the 1,400~s F656N WFPC2 image using the ``polyphot'' routine 
in IRAF.  From the method for narrow-band photometry 
outlined on the WFPC2 
website\footnote{http://www.stsci.edu/instruments/wfpc2/Wfpc2\_faq/wfpc2\_nrw\_phot\_faq.html}, 
we have determined the \ha\ flux to be 
\mbox{1.5 $\times$ 10$^{-14}$ erg s$^{-1}$ cm$^{-2}$}.
Using the E(B$-$V) value of 0.52~mag and the distance value 
of 5.1~Mpc, we find a total \ha\ luminosity of
\mbox{1.4 $\times$ 10$^{38}$ erg s$^{-1}$}.  
This value is consistent with the
luminosity reported by \citet{BF94},
\mbox{1.9 $\times$ 10$^{38}$ erg s$^{-1}$}, within 
their error limit of 25\%.
Under the assumption that the broad line component 
originates from the expanding SNR shell, the ratio
between the broad and narrow components suggests that
\mbox{$\sim$8.5 $\times$ 10$^{37}$ erg s$^{-1}$}
of the total \ha\ luminosity originates from
the expanding SNR shell.  We have used this 
broad component \ha\ luminosity to determine the 
mass and kinetic energy of the SNR shell.  We have 
assumed an ellipsoidal shell structure with a 
third axis of 25~pc and a shell thickness of 1\% of 
the radius.  We have also used the simple assumption 
that the remnant is composed of singly ionized 
hydrogen and helium.  Using the electron density 
determined by \citet{BFS99}, \mbox{410$\pm$100 cm$^{-3}$},
we find a filling factor of 0.21$^{+.15}_{-.08}$, a mass of 
\mbox{640$^{+200}_{-130}$ M$_{\odot}$} and a kinetic energy of 
\mbox{3.2$^{+1.0}_{-0.7}$ $\times$ 10$^{50}$ erg} for the SNR 
shell (See Table~\ref{tblphysical}).

\subsection{X-Ray Properties}
\label{sec:xray}

\begin{figure}
\epsscale{1}
\plotone{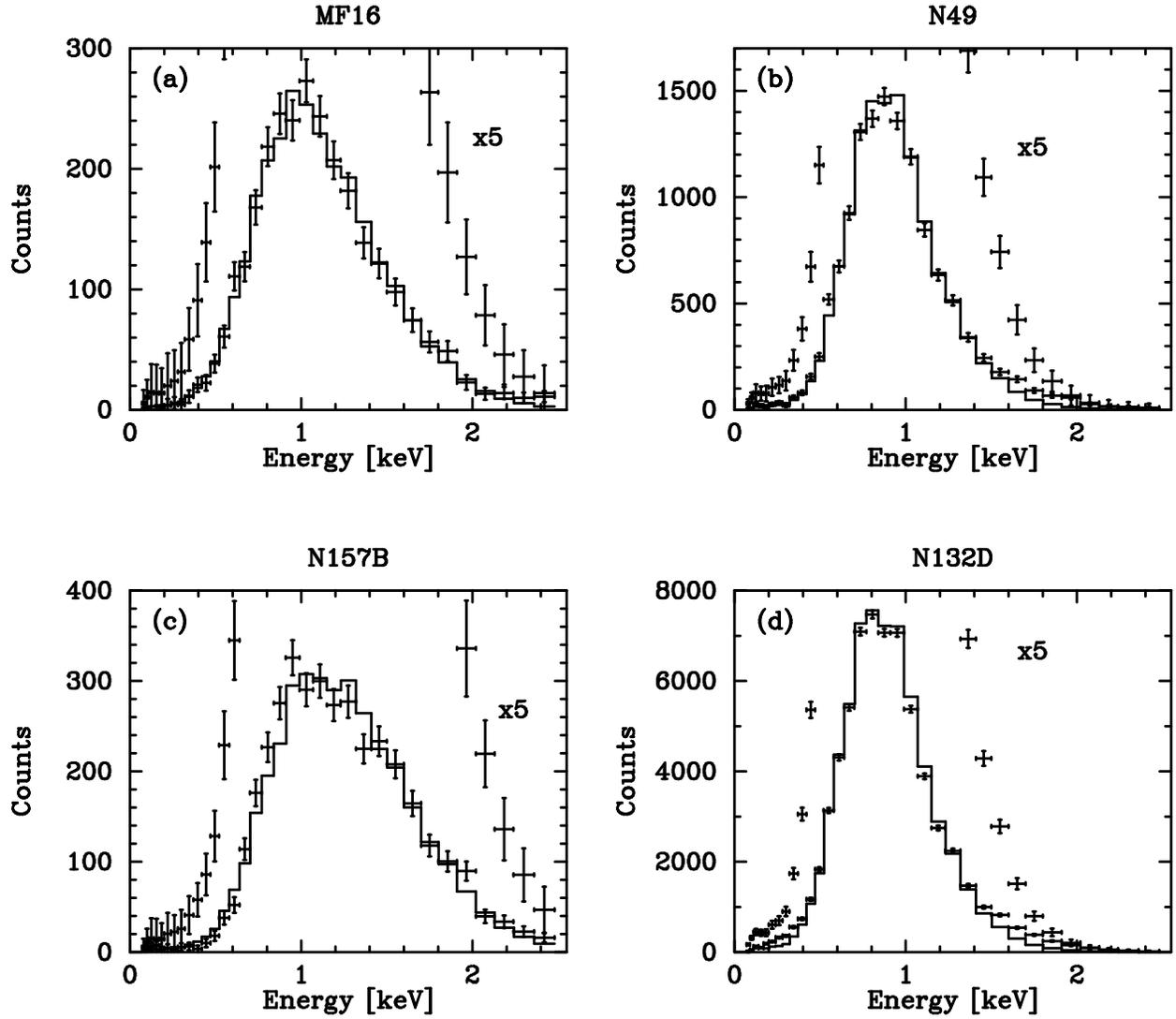}
\caption{{\it ROSAT} PSPC X-Ray spectra for (a) MF16, 
(b) N49, (c) N157B, and (d) N132D.  In each panel, the spectrum is
plotted twice, with the upper plot exaggerated by a factor of 5 to
emphasize the high-energy tail above 2~keV.  The lower plot shows
the best-fit model: power-law model for MF16 and N157B, Raymond \& Smith's
(1977) thin plasma emission model for N49 and N132D. \label{figxray}}
\end{figure}

We have used the {\it ROSAT} PSPC observation of NGC\,6946 to 
extract the X-ray spectrum of MF16.  To fit the X-ray spectrum, 
we applied both a thin plasma emission model \citep{RS77} and 
a power-law model.  We found a power-law model with energy index 
\mbox{$\alpha$$=$3.7$\pm$0.6} and absorption column density 
\mbox{N$_{{\rm H}}$$=$5.6$^{+1.5}_{-1.1}$ $\times$ 10$^{21}$ cm$^{-2}$} 
to be the best fit for the {\it ROSAT} data (See 
Figure~\ref{figxray}a) with \mbox{min($\chi$$^{2}$) = 25.5},
half that of the best thin plasma emission model fit.  
Our power-law fit is consistent with that of \citet{S94}, who concluded
that the X-ray spectrum of MF16 could be fitted equally
well by several simple models, not including the Raymond-Smith 
thin plasma emission model.  The ambiguity in the model fit was
attributed to the low spectral resolution and the very soft
energy band of the PSPC.
We note that the PSPC spectrum of MF16 has a high-energy tail
above 2~keV, as shown in Figure~\ref{figxray}.  This
high-energy emission has been confirmed by an ASCA observation,
which shows a complex spectrum that requires multiple emission
components \citep{SB98}.

\subsection{M33 Remnants}
\label{sec:m33}

We have examined the echelle spectra of the M33 SNRs
by the same methods used for the echelle spectra of MF16.  
The broad line components of the M33 remnants show 
\nii$\lambda$6584/\ha\ ratios of 0.3--0.55.  The narrow line components 
show \nii$\lambda$6584/\ha\ ratios of 0.15--0.35, similar to the \hii\ 
regions near the SNRs.  A complete summary of the M33 SNRs 
can be found in Table~\ref{tblm33}.  These \nii$\lambda$6584/\ha\ ratios are
significantly lower than the \nii$\lambda$6584/\ha\ ratios of the remnant 
MF16.

Even the remnants with velocity widths similar to those of
the SNR MF16 have weaker \nii$\lambda$6584/\ha\ ratios than those in MF16.  
M33-6, M33-8, M33-9 have 
\ha\ broad line component FWHM $\sim$270~\kms\ and \ha\ narrow 
line component FWHM $\sim$40~\kms, very close to the 
\ha\ velocity widths of MF16.  The broad line 
component \nii$\lambda$6584/\ha\ ratios for the M33-6, M33-8, and 
M33-9 are, however, only half as strong 
as those in MF16, and the narrow line component
\nii$\lambda$6584/\ha\ ratios are only one-third as strong as 
those in the ultraluminous SNR MF16.

\section{Discussion}
\label{sec:discuss}

\subsection{Luminosity Consideration}
\label{sec:luminosity}

The SNR MF16 has an extraordinary luminosity;
it is comparable in optical brightness to the young, ultraluminous
SNR in NGC\,4449 and $\sim$3.5 times brighter in X-rays.  
Despite this, the supernova does not appear to 
have been highly energetic.  The kinetic energy of the 
remnant shell, \mbox{$\sim$3 $\times$ 10$^{50}$ erg},
is similar to the shell kinetic energies of 
\mbox{$\sim$10$^{50}$ erg} determined by \citet{Rosa97,Rosa99a} 
for remnants in the Magellanic Clouds, ruling out the 
necessity of a hypernova remnant \citep{Wang99} or other exotic 
object with explosion energy $\gtrsim$ 10$^{53}$~erg.  
The remnant does not appear 
very young (\mbox{$\sim$10$^{2}$ yr}) because it has a low 
expansion velocity and a moderate size.  
If we assume that this SNR is formed by a single supernova
in a homogeneous medium and that the SNR is in an 
adiabatic expansion stage, we may
use the \citet{Sedov} solution to derive an age of $\sim$25,000 yr.  
This age is an order of magnitude greater than the upper age
limit determined by \citet{BF94}, $\leq$3,500 yr.  
While we make no claim the Sedov solution
accurately describes MF16, it demonstrates that the remnant 
could be millennia in age.  

If MF16 is not highly energetic or very young, 
then we must find another explanation 
to generate the luminosity of the remnant.  A possible mechanism
for the luminosity could be the interaction of the 
expanding SNR shell with dense material around the SNR.  
The young, ultraluminous SNR in NGC\,4449, which has 
optical luminosity comparable to the SNR MF16
\citep{BF94}, is luminous 
because it is expanding into a nearby \hii\ region 
\citep{BKW83}.  MF16 may also be luminous
because it is expanding into dense nearby material.
As emissivity $\propto$ N$^{2}_{e}$V, where N$_{e}$
is the electron density and V is the emitting volume,
a moderate increase in the density can greatly increase 
the luminosity.  We note that the existence of a bright 
narrow line component in MF16 indicates the existence of 
such dense circumstellar material.  

The interaction of MF16's SNR shell with dense circumstellar 
material can explain the optical luminosity, but it is
not clear whether or not this interaction can produce the 
high X-ray luminosity.  To further explore 
the X-ray properties of MF16, we have compared 
the {\it ROSAT} PSPC observations of this remnant 
to those of the LMC SNRs N49, N132D, and N157B from \citet{Rosa99b}.  
In Figure~\ref{figxray}, we compare MF16 to 
the LMC SNRs.  These three LMC SNRs have been previously
observed by the {\it Einstein} Solid State Spectrometer.
From these {\it Einstein} data, \citet{C82} find
that the X-ray spectra of N49 and N132D are best described 
by a thin plasma emission model and that the spectrum of 
N157B is best described by a power-law model.  In \S~\ref{sec:xray},
we determined that MF16's X-ray spectrum is best fitted by a 
power-law model and shows significant emission above 2~keV.  
The spectral characteristics are most similar to those of N157B.
Based on its radio and X-ray spectra, N157B has been 
identified as a Crab-type SNR \citep{Mills78,C82,Mat83}, and it 
is known to contain a pulsar \citep{Mar98}.  Based on the spectral 
similarities between MF16 and N157B, we suspect that a 
significant portion of the X-ray luminosity of MF16 is
contaminated by a non-thermal-plasma source.  To determine the
nature of the non-thermal-plasma source, we need high spatial resolution
and high spectral resolution X-ray observations combined with
a sensitive timing experiment.  At present, it is premature to 
determine the real fraction of thermal plasma emission in the
X-ray spectrum of MF16.

\subsection{Significance of the \nii$\lambda$6584/\ha\ Ratio}
\label{sec:nii/ha}

Much can be learned from the \nii$\lambda$6584/\ha\ ratios of 
the broad and narrow line components of an SNR's optical 
spectrum by comparing them to those of nearby \hii\ regions.  We 
use the SNRs of M33 to illustrate this effect.  As described 
in \S~\ref{sec:m33}, most of the observed SNRs in M33 are 
adjacent to \hii\ regions, and the observed \nii$\lambda$6584/\ha\ 
ratios in the narrow components at the SNRs are similar to those 
in the adjacent \hii\ regions.  This indicates that the narrow 
components originate in the un-shocked material surrounding the SNRs.  
The broad components of the SNRs are associated with the expanding
shells of shocked gas.  As the shocked gas has higher temperatures, 
and as forbidden line strengths increase steeply with temperature, 
the forbidden/recombination line ratios should be higher in shocked 
gas with higher temperatures.  Indeed, the M33 SNRs show higher
\nii$\lambda$6584/\ha\ ratios in the broad components than in the narrow 
components.

The SNR MF16 has unusually high \nii$\lambda$6584/\ha\ ratios.  
The total \nii$\lambda$6584/\ha\ ratio of MF16 is roughly 
twice the average \nii$\lambda$6584/\ha\ 
ratio for other SNR candidates in NGC\,6946, $\sim$0.45 \citep{MF97}.
Both its broad line component and the narrow line component have 
higher \nii$\lambda$6584/\ha\ ratios than the observed M33 SNRs.  
We do not have echellograms of any nearby \hii\ regions for a direct
comparison; however, several \hii\ regions at galactocentric 
distances of 2.5~kpc and 4.5~kpc, similar to that of MF16 
(3.7~kpc), have been observed by \citet{Mc85} and can be compared 
with the narrow line component of the remnant.  These \hii\ regions
in NGC\,6946 have \nii$\lambda$6584/\ha\ ratios $\sim$0.3, similar to 
those seen in \hii\ regions and SNR narrow line components in
M33, but are much lower than that of
the narrow line component of the NGC\,6946 remnant, 0.8.

High \nii/\ha\ ratios can be associated with shock excitation
or an enhanced nitrogen abundance.  While the shock excitation may be
partially responsible for the high \nii$\lambda$6584/\ha\ ratios in the
broad line component, shock excitation cannot be responsible for
the narrow line component.  The FWHM of the narrow component of the
\nii$\lambda$6584 line is only 25~\kms\ for the SNR MF16, implying a 
gas motion of $\sim$13~\kms, which is only slightly larger 
than the isothermal sound velocity of a 10$^4$~K gas.  We are left 
with the inevitable alternative that an enhanced nitrogen abundance 
causes the high \nii$\lambda$6584/\ha\ ratio in the narrow line component 
of the remnant.

Elevated nitrogen abundance and high \nii/\ha\ ratios are a
hallmark of stellar ejecta nebulae associated with 
nitrogen-enhanced Wolf-Rayet (WN) stars and luminous blue 
variables (LBVs) (e.g., Esteban et al. 1992; Smith et al. 1998).  
The SNR MF16 may be interacting with 
a circumstellar nebula produced by the supernova's progenitor.  The
expansion velocity of this nebula, $\sim$13~\kms, implied by 
the FWHM of the narrow \nii$\lambda$6584 component, 
is near the low end of the range of expansion velocities 
observed in ring nebulae around massive stars \citep{CWG99}.  

SNRs interacting with the progenitors' ejecta nebulae can be observed
only when the SNRs are young, as old SNRs' evolution will be dominated
by the distribution of ambient interstellar medium.  This interaction 
is actually commonly seen in young SNRs.  The two youngest known SNRs in 
the LMC, SN1987A and SNR 0540$-$69.3, both show a circumstellar
nebula with an enhanced \nii/\ha\ ratio \citep{Burr95,CMB98}.  
The youngest SNR in our Galaxy,
Cas A, also shows traces of its progenitor's ejecta.  Cas A contains
bright, slow-moving knots of gas called quasi-stationary flocculi (QSF).
The QSFs have \nii/\ha\ ratios of 1--4, and nitrogen abundances 
enhanced by a factor of 10 \citep{KC77}.  The QSFs have
been suggested to be ejected by a WN progenitor of the supernova 
\citep{PvB71,CK78}.  We propose that MF16 is a young SNR, 
$\lesssim$10$^{3}$~yr, created by the supernova explosion 
of a WN or LBV star, and that it is interacting with the 
ring nebula produced by the progenitor.

\subsection{Colliding SNRs vs. Single SNR}
\label{sec:vs}

\begin{figure}
\epsscale{0.60}
\plotone{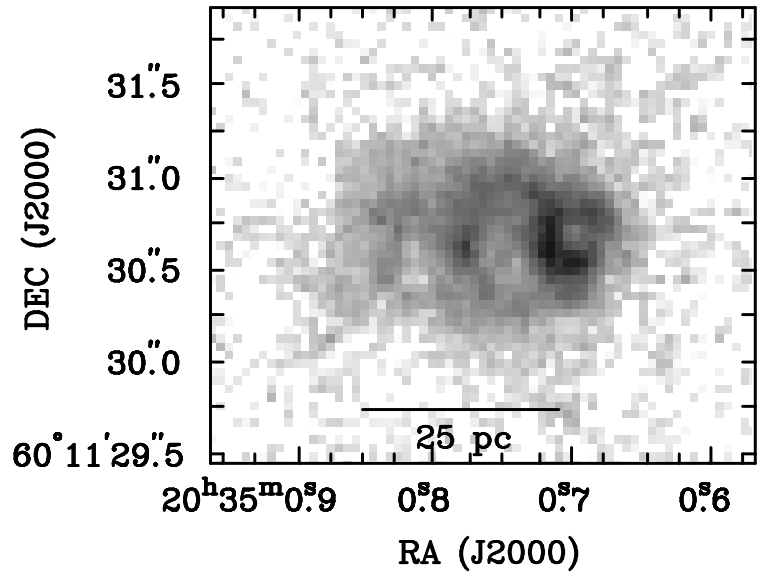}
\caption{Narrow field-of-view WFPC2 F656N (\ha) image 
of MF16 to show multiple-loop morphology. \label{figHSTHa}}
\end{figure}

It has been suggested that MF16 is the product of 
colliding SNRs of different ages \citep{BFS99}.
This colliding SNR hypothesis is based on the 
multiple-loop morphology seen in the F656N WFPC2 image
(See Figure~\ref{figHSTHa}).  
The colliding SNR hypothesis requires the 
detonation of at least two massive stars in close 
proximity to one another in both space and time.  
\citet{BFS99} support this possibility by
suggesting that MF16 is embedded in an OB 
association.  As discussed in \S~\ref{sec:environment},
any association in this region would be 
$\gtrsim$10$^{7}$~yr old.  We would therefore expect
the O--early-B type stars of this association to have 
``burned out'' already.  This is also the disintegration 
timescale of OB associations \citep{SE88}.  Thus,
any association in the region is comprised of mid- to late-B
type stars and is mixing into the general stellar population.
We suggest that the probability of multiple supernovae 
occurring in this environment, with the proper spatial
and temporal proximity to one another, is very low.

The multiple-loop morphology of MF16 is also consistent
with a single SNR hypothesis.  During the evolution
of a massive star, the main-sequence winds, stellar ejecta, 
supergiant winds, and/or Wolf-Rayet winds shape multiple 
shells of ambient material around the evolving star 
\citep{Marston95,GS96a,GS96b}.  This effect is clearly 
seen in the nebula NGC\,6888, which contains a WN star 
surrounded by a nitrogen-enriched circumstellar bubble of 
of size 12$\times$6 pc and an interstellar bubble of diameter 28 pc
\citep{Cappa96}.  As a massive star inside such a nebula 
explodes as a supernova, the resultant SNR will have a 
multiple-loop morphology.  We therefore 
suggest that a single SNR hypothesis is favored
over a colliding SNR hypothesis for MF16.

\section{Conclusions}
\label{sec:conclusions}

We conclude that the ultraluminous SNR MF16 in NGC\,6946 is
a normal SNR expanding into a complex ring nebula formed by 
a massive progenitor, specifically a WN star or LBV.  Such 
a nebula is expected to contain a fragmented circumstellar 
bubble of stellar ejecta surrounded by a larger interstellar 
bubble of ambient material swept up by stellar winds 
\citep{GS96a,GS96b}.  The optical luminosity of the remnant 
is caused by the interaction of the SNR shell with these 
dense shells of material in the nebula.  The multiple-loop 
morphology seen in the F656N WFPC2 image is a result of this 
interaction.  Since the supernova exploded into the cavity 
of this nebula, the SNR shell would have expanded rapidly 
through the fragmented circumstellar bubble until it 
encountered the outer interstellar bubble.  The remnant is 
therefore still young, $\lesssim$10$^{3}$ yr, and the 
interior of the SNR contains dense clumps of stellar ejecta 
being ablated by the SNR shock and evaporated by the hot 
medium of the SNR interior.  We further suggest that 
because the progenitor was either a WN star or an LBV, its 
circumstellar material is nitrogen-rich, leading to the 
strong \nii/\ha\ ratio.  Similar high luminosity is expected 
when the ejecta of SN1987A hits the circumstellar rings, which 
have strong \nii/\ha\ ratios \citep{Burr95}.  Future observations 
with a spectroscopic instrument able to resolve the nebula,
such as the Space Telescope Imaging Spectrograph, would allow 
the spectral properties of individual sections of the SNR MF16 
to be probed, resolving many of the outstanding questions of 
this most interesting supernova remnant.

\acknowledgments{We would like to thank William Blair for 
his useful communications in preparing this paper and for being 
a co-observer on the M33 echelle data.  We also thank 
Schuyler Van Dyk for being a co-observer on the MF16 echelle 
data, Rosie Chen for valuable discussions in analyzing the 
WFPC2 data, and Rosa Williams for access to her X-ray atlas
of LMC SNR.  Finally, we thank the anonymous referee for 
his/her helpful comments.  This research is supported by 
NASA grant NAG 5-7003.}

\clearpage

\clearpage

\clearpage

\begin{table}
\caption{MF16 Spectral Velocity Widths \label{tblwidth}}
\begin{tabular}{l c c}
\tableline
\tableline
  & Broad & Narrow \\
  & Component & Component \\
\tableline
\ha\ FWHM (\kms) & 285 & 42 \\
\nii$\lambda$6584 FWHM (\kms) & 250 & 25 \\
\oiii$\lambda$5007 FWHM (\kms) & -- & 22 \\
\ha\ FWZI (\kms) & 450 & -- \\
\nii$\lambda$6584 FWZI (\kms) & 400 & -- \\
\oiii$\lambda$5007 FWZI (\kms) & 430 & -- \\
\tableline
\end{tabular}
\end{table}


\begin{table}
\caption{MF16 Comparative Flux Levels \label{tblflux}}
\begin{tabular}{l c}
\tableline
\tableline
  & Ratio \\
\tableline
\nii$\lambda$6584/\ha\ broad & 1.0 \\
\nii$\lambda$6584/\ha\ narrow & 0.8 \\
\ha\ broad/narrow & 1.5 \\
\nii$\lambda$6584 broad/narrow & 1.8 \\
\oiii$\lambda$5007 broad/narrow & $\sim$1 \\
\tableline
\end{tabular}
\end{table}


\begin{table}
\caption{Physical Parameters Of MF16's SNR Shell \label{tblphysical}}
\begin{tabular}{l c}
\tableline
\tableline
Parameter & Value \\
\tableline
Luminosity(\ha) & 1.4 $\times$ 10$^{38}$ erg s$^{-1}$ \\
Electron Density\tablenotemark{a} & 410 cm$^{-3}$ \\
Filling Factor & 0.21 \\
Mass & 640 M$_{\odot}$ \\
Kinetic Energy & 3.2 $\times$ 10$^{50}$ erg \\
\tableline
\end{tabular}
\tablenotetext{a}{From \citet{BFS99}}
\end{table}

\clearpage

\begin{table}
\caption{M33 SNRs Flux Ratios \& Spectral Velocity Widths \label{tblm33}}
\begin{tabular}{l c c c|c c c|c}
\tableline
\tableline
  & \multicolumn{3}{c}{Broad} & \multicolumn{3}{|c|}{Narrow} & \hii\ \\
  & \multicolumn{3}{c}{Component} & \multicolumn{3}{|c|}{Component} & Region \\
\tableline
  & \nii\tablenotemark{a}/\ha\ & \multicolumn{2}{c|}{FWHM} & 
\nii\tablenotemark{a}/\ha\ & \multicolumn{2}{c|}{FWHM} & 
\nii\tablenotemark{a}/\ha\ \\
  &  & \ha\ & \nii\tablenotemark{a} &  & \ha\ & \nii\tablenotemark{a} & \\
  &  & (\kms) & (\kms) &  & (\kms) & (\kms) & \\
\tableline
M33-2  & 0.3  & 190 & 190 & 0.15 & 45 & 30 & 0.15 \\
M33-4  & 0.3  & 140 & 145 & 0.2  & 40 & 30 & 0.3  \\
M33-6  & 0.5  & 255 & 240 & 0.25 & 40 & 25 & 0.25 \\
M33-8  & 0.5  & 275 & 265 & 0.3  & 40 & 40 & 0.3  \\
M33-9  & 0.45 & 270 & 255 & 0.25 & 40 & 40 & 0.2  \\
M33-11 & 0.55 & 140 & 150 & 0.35 & 50 & 50 & 0.25 \\
M33-16 & --   & --  & --  & 0.2  & 60 & 50 & 0.2  \\
M33-18 & 0.4  & 145 & 140 & 0.25 & 30 & 25 & 0.15 \\
\tableline
\end{tabular}
\tablenotetext{a}{\nii$\lambda$6584}
\end{table}

\end{document}